\documentclass[aps,prc,onecolumn,showpacs,floatfix,groupedaddres]{revtex4}

\usepackage{graphicx}

\begin{document}

\title {\bf The Single Particle Sum Rules in the Nuclear Deep-Inelastic Region}

\author {J.Ro\.zynek and G.Wilk}

\affiliation{The Andrzej So\l tan Institute for Nuclear
Studies, Ho\.za 69; 00-689 Warsaw, Poland}

\date{\today}

\begin{abstract}
We have modelled the parton distribution in nuclei using a
suitably modified nuclear Fermi motion. The modifications concern
the nucleon rest energy which changes the  Bj\"orken $x$  in a
nuclear medium. We also introduce final state interactions between
the scattered nucleon and the rest of the nucleus. The
energy-momentum sum rule is saturated. Good agreement with
experimental data of the EMC effect for $x> 0.15$ and nuclear
lepton pair production data has been obtained.
\end{abstract}


\maketitle

Deep inelastic scattering of electrons on nuclei (NDIS) proceeding
with a highly virtual photon, $|Q^{2}|>1$ GeV, provides us with a
picture of bound nucleons with partially deconfined constituents
due to the presence of the surrounding nuclear matter
\cite{MEZONS}. The resolution $1/\sqrt{|Q^{2}|}$ is good enough to
allow for the space localization of nucleonic constituents. The
influence of a nuclear medium on the nucleon structure function
(SF), $F_2(x)$, (known as EMC effect) also depends on the
longitudinal scale, $z=1/(M_Nx)$ \cite{Foot1}, on which it is
observed and which is determined by the mean free path and/or
lifetimes of the struck partons \cite{Ja}. For small $x$, the
corresponding scale $z$  exceeds the mean nucleon-nucleon ($NN$)
distance in the nuclear matter. Therefore, in this region, the EMC
effect should be described using some collective degrees of
freedom characterized by the corresponding size of the nuclear
area and by nucleonic (or partonic) clusters. On the other hand,
for $x>x_B=0.6$ one has $z<0.3$ fm, i.e. the lifetime of the hit
parton is sufficiently small for its final state not to be
affected by the $NN$ interaction because it travels a distance
comparable with the nucleon hard core radius, $r_C\simeq0.35$ fm,
much smaller than nucleon average radius, $r_N\simeq 0.8$ fm. It
follows that in this region the single nucleon degrees of freedom
play a crucial role in the description of the nuclear influence on
the partonic SF of nucleon.

\begin{figure}[h]
\includegraphics[width=7.1cm,height=43mm]{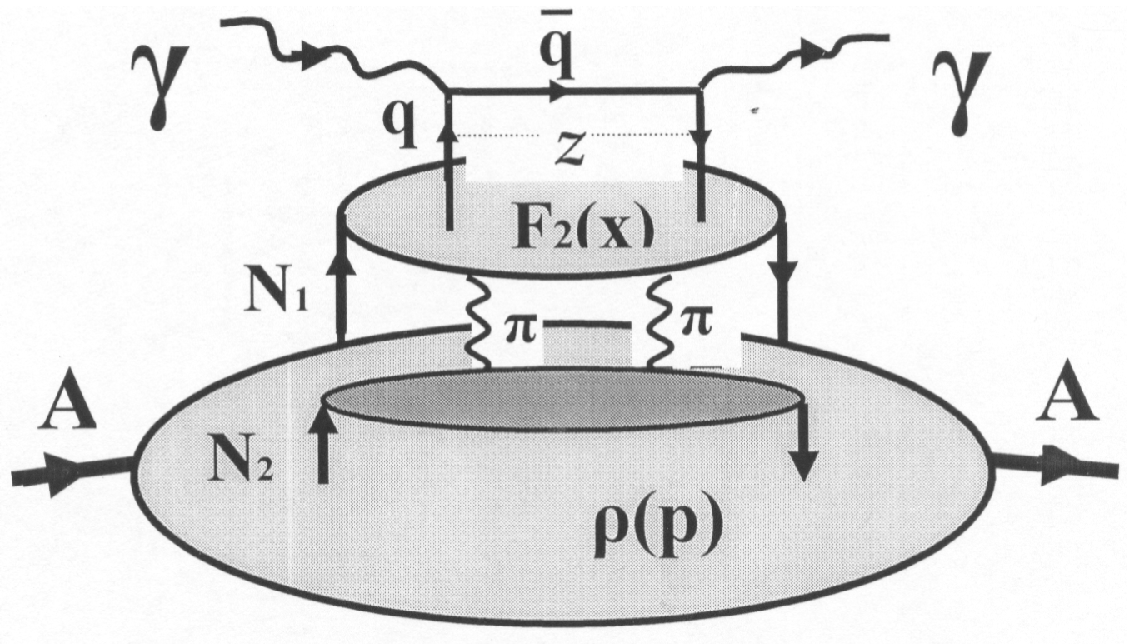}
\caption{The graphical representation of the convolution model for
 the deep inelastic electron nucleus scattering
with 2$\pi$ Final State  Interaction (FSI) between the active
nucleon $N_1$ (with hit quark $q$) and the spectator $N_2$.}
\label{fig:Figure1}
\end{figure}

In this work, we investigate the region $0.15<x<1$. Following
\cite{Ja}, a usual two step mechanism of NDIS is adopted,
accounting for the fact that: $(a)$ nuclei are composed of $A$
nucleons, which are distributed according to the nuclear
distribution function, $\rho^{A}(y_A)$); $(b)$ nucleons are
composed of partons, which are described by the free nucleon
structure function (SF) $F_2^{N}(x)$). The nuclear SF is therefore
the following convolution (see Fig.\ref{fig:Figure1})
\cite{Foot2}:
\begin{equation}
\frac{1}{x_{A}}F^{A}_{2}(x_{A}) = A \int \int dy_{A} \frac{dx}{x}
\delta (x_{A}-y_{A}x) \rho ^{A}(y_{A}) F^{N}_{2}(x) .
\label{eq:comp}
\end{equation}

This simple formula failed as a good description of nuclear SF.
Some additional degrees of freedom present only in nuclear matter,
usually attributed to additional mesonic \cite{MEZONS} could be
present. However, mesons alone cannot account for data on lepton
pair production on nuclei \cite{Alde}. It suggest that, the
nucleon SF $F_2^{N}$ itself is modified in the medium. Both
effects should be included.

In this paper, we shall  examine the following picture. At large
values of $x$, $x\in (x_B,1)$, the longitudinal resolution in $z$
is small enough and NDIS proceeds only on a single nucleon but
with its mass modified accordingly by the presence of the nuclear
medium. On the other hand, approaching $x=0.15$ limit, the
longitudinal scale gradually grows, eventually exceeding nucleonic
size and later also inter-nucleonic distances in the nucleus. The
action of other nucleons is now strong and should be accounted
for. Because the approximation of independent nucleonic pairs is
already responsible for $\sim80$\% of nuclear binding, we shall
consider in this region NDIS as proceeding on such  an interacting
pair which form quasi-bound state (cf. Fig. \ref{fig:Figure1}).
This effect will be modelled by an $x$ dependent mass of the
nucleon \cite{Foot3}. However, to satisfy exactly the momentum sum
rule (MSR), one has to allow for a nuclear pion excess, which
should be small (of the order of $\sim 1\%$) in order to
simultaneously describe the lepton-pair production data.

Let us start with the $x>x_B$ regime, where the partonic mean free
paths $z$ are much shorter then the average distances between
nucleons. It means that we can treat nuclear nucleons as
noninteracting objects remaining on the energy shell and not
affected by FSI shown in Fig.1:
\begin{equation}
\sqrt{p^2} \equiv {M_B} _{\overrightarrow{{~rest~}}} p^+ .
\label{eq:onshell}
\end{equation}
Let $M_B$, identified with the $p^+$ component in the nucleon rest
frame, denote the rest energy of the nucleon in this case. To
calculate it, we assume that nuclear longitudinal momentum
$P_A^{+}$ component is given as a sum of all partonic momenta
$k_{Ai}^+$. For $A$ nucleons we have
\begin{equation}
\frac{1}{A} \sum_{i=1}^{nA} k_{Ai}^+ = \frac{M_A}{A} \equiv M_N +
\epsilon = \int d^{3}p \sqrt{{M_B}^2 + \vec{p}^2}  ,
\label{eq:estim}
\end{equation}
where $\epsilon \simeq -8$ MeV is the usual nucleon binding energy
and $n$ is the number of partons inside the nucleon (in what
follows we shall work with $A>50$ nuclei and assume uniform
momentum distribution of nucleons).

For the uniform nucleon Fermi
distribution the average nucleon Fermi energy is
\begin{equation}
E_{Fermi}\simeq 0.6 \cdot \left(\vec{p}_F^2/(M_N+\epsilon)\right)
\label{eq:Fermi}
\end{equation}

\noindent
where ($p_F =(3\pi^2\bar{n}_A/2)^{1/3}$ is average Fermi momentum given
by average nuclear density $\bar{n}_A$). One gets from
(\ref{eq:estim}) that:
\begin{equation}
M_B \cong M_N + \epsilon - E_{Fermi}.\label{eq:mamam}
\end{equation}
It means that in a nuclear medium characterized by $\epsilon$ and
$E_{Fermi}$, the rest energy  of the nucleon, $M_B=\sum_i
k_{Ni}^+$, takes in the large $x$ limit, $x> x_B$, a value
different from the sum of the corresponding partonic energies
$k_{Ni}^0$ expressed in the rest frame of the nucleon (notice that
they differ from $k_{Ai}^0$ in (\ref{eq:estim})); here and in what
follows we put $x_B=0.6$ because, as was stated before, this value
corresponds to longitudinal scale $z$ starting to be smaller than
the nucleon hard core radius). The $M_B < M_N$ represents
therefore the additional effect of Fermi motion emerging from the
partonic ($x$) level of description.

Following the line of reasoning proposed by us in \cite{RW03}, we
are not changing the expression for the nuclear spectral function
$\rho^A(y_A)$ and take it from the Relativistic Mean Field (RMF)
model of the nucleus \cite{SW}. In the relativistic Fermi gas
model \cite{B} it is given by:
\begin{eqnarray}
\rho ^{A}(y_A) =&& {4\over \rho }\int {d^{4}p\over (2\pi)^{4}}
S_N(p^o,{\bf p})\nonumber\\
&&  \left[1+\frac{p_3}{E(p)}\right] \delta \left[y -
\frac{(p_o+p_3)}{\mu}\right] , \label{eq:final}
\end{eqnarray}
where the factor $(1+p_3/ E(p))$ representing relativistic
correction \cite{FS} has been obtained for the RMF form of the
nucleon spectral function: $S_N=n(p)\delta ( p^0-(E(p) +U_V))$
with $E(p)=\sqrt{(M_N+U_S)^2+{\bf p}^2}$.

Let us now proceed to lower values of $x$, $x<x_B$. This is the
region in which an increase of the partonic mean free paths makes
them eventually comparable with internucleonic distances and one
faces the problem of how to properly treat forces binding nucleons
in nuclei. To solve it, notice that for sufficiently small values
of $x$, in the region of $x<x_{N}\simeq 0.3$, the uncertainties in
the lifetime of an intermediate parton state are so big that one
should include exchanges of nuclear mesons (like two $\pi$, but
also $\sigma$, $\omega$ and $\rho$) between nucleons, cf. Fig.
\ref{fig:Figure1}. In standard low energy nuclear physics this is
usually done by adding to equations (\ref{eq:estim}) and
(\ref{eq:mamam}) the nuclear potential energy term resetting the
effective nucleon mass to its free nucleon value $M_N$. To be more
specific, for $x>x_B=0.6$ we have $z(x)=1/(xM_N) <r_{C} = 0.35$
fm. In this region, the nearby second nucleon, which is separated
by the average distance $r_{h} \simeq 1.7$ fm (obtained from
independent pair approximation) , will not affect collisions
proceeding on the active nucleon. The situation changes when $x$
is getting smaller and the uncertainty $z$ exceeds the nucleon
radius $r_N=0.85$ fm $ \simeq r_h/2$. For such an $x$, the single
nucleon approximation is no more applicable. We shall model this
process by introducing probability $f(x)$ that the struck quark
originates from the nucleon which interacts with a nearby nucleon
spectator and assume for its simple linear form in the
$z=(1/(M_Nx)$ variable (i.e. in the longitudinal scale):
\begin{equation} f(x)=\left\{ \begin{array}{lcl}
        0 & {\rm if} & x \geq x_B  \quad (\simeq 0.6)\\
        \left[\frac{1}{M_Nx}-r_c\right]/(r_{N}-r_c)   \\
        1 & {\rm if}&  x\leq \frac{1}{r_{N}M_N} \quad (\simeq 0.25)
        \end{array} \right .
 \label{int}
\end{equation}
This function  determines whether interaction proceeds on the
noninteracting nucleon with the rest energy equal to $M_B$ or on
the correlated nucleon with a different rest energy. One can
introduce the effective nucleon SF, $\tilde{F}^N_2$, and express
it via one- and two-nucleonic SFs:
\begin{equation}
\tilde{F}^{N}_{2}(x)= f(x) F^{2N}_{2}(x)+ (1-f(x))F^{N}_{2}(x),
\label{eq:distribution}
\end{equation}

Now we can take into account the effect of two body FSI in
$F^{2N}_{2}(x)$ by changing the nucleon rest energy $M_B$ and
adding the $NN$ interaction term to the equation (\ref{eq:mamam}).
Assuming the dominant role of two-body short range $NN$
correlation, it will reset nucleon the rest energy to the standard
low energy value $M_N$. It gives the dominant change of the
nucleon SF for smaller $x<x_B$ coming from FSI. We are not
including other effects of the $NN$ interaction.

Now, the above interpolation procedure applies also approximately
to the rest energy of the struck nucleon. Using the relation
$<V_N> = 2(\epsilon - E_{Fermi})$, which connects the average
nucleon interacting energy $<V_N>$ with the Fermi momentum in the
nucleus, the interesting nucleon rest energy now becomes
$x$-dependent and is given by \cite{Foot4}:
\begin{equation}
M_x = M_N + \frac{(1-f(x))}{2}<V_N> \label{eq:masses}
\end{equation}
Such an $x$-dependent nucleonic mass affects the MSR which is now
given by:
\begin{eqnarray}
\frac{\frac{1}{A}\int F^{A}_{2} (x_A) dx}{\int F^{N}_{2} (x) dx}\,
&=&\, \left( 1-
C_x\frac{M_N-M_B}{M_N} \right)\frac{M_A}{AM_N}\, =\nonumber\\
&=&\, \left( 1- C_x\frac{E_{Fermi} - \epsilon}{M_N}
\right)(1+\epsilon) , \label{eq:estimation}
\end{eqnarray}
where the factor $C_x$  comes from the integration of the function
f(x) and indicates the amount of momentum  violation (no violation
for $C_x =0$ and maximum violation for $C_x=1$). For constant mass
$M_x=M_B$ one has $C_x=1$ in the whole range of $x$ and
consequently this sum rule is violated by $\sim 3$\% \cite{Foot5}.
The increase of nucleon rest energy connected with the inclusion
of nuclear FSI in NDIS  decreases $C_x$ and will reduce  this
violation. Calculations performed for $^{56}Fe$ using
$\bar{n}_A=0.12$ fm$^{-3}$, $\epsilon = -8.8$ MeV and with
$x$-dependent nucleon mass (\ref{eq:masses}) reduces this
violation further to the level of $1\%$ only. This amount can be
safely (i.e. in agreement with nuclear lepton pair production data
\cite{DY}) accounted for by increasing the sea quarks momentum
fraction in the nuclear medium. It can be done by the scaling of
the $x$ variable  by 3\% in the sea quark part of SF
${F}^{N}_2(x)$ and it corresponds in model \cite{RW03,EI} to the
similar increase of the mass of virtual pion for $Q^2=4$ $GeV$
(for $Q^2=30$ $GeV$ it would be $<1$\%) \cite{Foot8} .

\begin{figure}
\includegraphics[width=12.1cm,height=75mm]{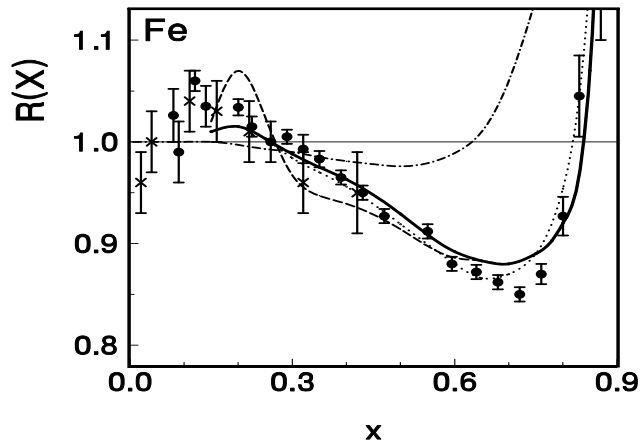}
\caption{Results and data \cite{Data} for the ratio
$R(x)=F^{56}_{2}(x)/F^{D}_{2}(x)$. $Fe$ data (solid circles) are
from SLAC. See also $Cu$ data from NA2' ($x$ marks). Curves
described in text.} \label{fig:Figure2}
\end{figure}

\begin{figure}
\includegraphics[width=12.1cm,height=75mm]{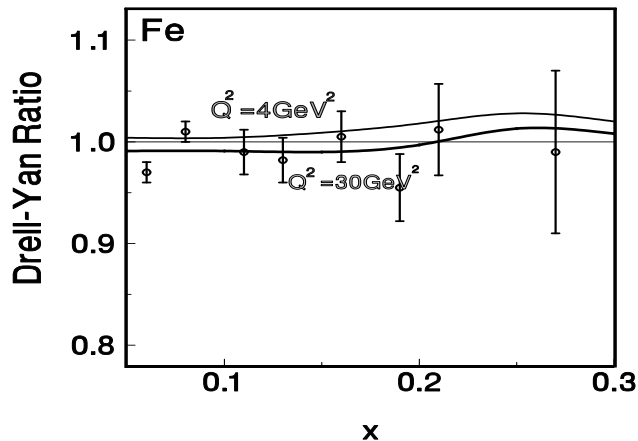}
\caption{Results for $Q^{2}=30$ GeV$^2$ and data \cite{Alde} for
the ratio of Drell-Yan lepton pair production. Results for
$Q^{2}=4$ GeV$^2$ (as in the EMC effect) are shown for reference.}
\label{fig:Figure3}
\end{figure}

The results of the calculations of the total nuclear SF, $R(x) =
{F}^{A}_2(x)/{F}^{D}_2(x)$, are presented in Fig.\ref{fig:Figure2}
\cite{FOOTA}. Notice that the case of pure "Fermi smearing", where
$M_x=M_N$ for the all $x$ (no binding), represented by the
dash-dotted line, fails completely. The situation improves
dramatically (cf. dashed line) when one uses a modified constant
nucleon rest energy, $M_x=M_B$ (i.e. $C_x=1$). In this case, to
satisfy the nuclear MSR, we have to enlarge the nuclear sea
contribution (pion excess) by allowing for $3\%$ of the momentum
to be carried by the nuclear sea quarks. However, this makes
agreement with lepton pair production data \cite{Alde} impossible.
Therefore in the next step (solid line) we have included the
$x$-dependent modification of the nucleon rest energy by using
$M_x$ as given by eq.(\ref{eq:masses}). In this case the nuclear
MSR is satisfied by increasing the momentum of sea quarks in bo
und nucleon by only $1$\%. Agreement with the data is now very
good \cite{Foot6}. The dotted line shows results obtained by
replacing $F_2^D$ by $F_2^N$, i.e. it shows effect caused by
deuteron.  As seen in Fig. \ref{fig:Figure3} the $1$\% increase of
the nuclear sea at  $Q^{2}=30$ is now fully compatible with the
lepton pair production data \cite{Alde}.

Consequently, although both results (represented by dashed and
solid lines) satisfy the MSR, one can see that they differ for
$x<0.6$ with the solid line presenting better fit to data in the
region of $0.3 < x <0.6$ (due to our $x$-dependent effective
nucleon mass). The sea quark contributions, which are located at
low $x$, are relatively small for the solid line and agree with
the data. Agreement with data for $x>0.6$ is determined only by
the values of $\epsilon$ and $p_F$ used in (\ref{eq:mamam}).
Finally, the overall fit represented by the solid curve is very
good \cite{Foot6}, better then the dashed one, and it shows that
the physical mechanism of the $x$-dependent nucleon rest energy in
NDIS works properly, although the results depend slightly on the
actual value of $x_B$. The presented model makes it possible to
incrementally switch on (depending on the scale $z$) the effect of
nucleon-nucleon interaction, giving the desired big reduction of
$C_x$ in the MSR (\ref{eq:estimation}).

To summarize: we propose to account for the EMC effect in NDIS by
using the single particle approach  with effective Fermi motion
caused by nuclear interactions, which is also responsible for
medium changes of the parton distributions inside nucleons. The
parton distribution is the function of the nucleon rest energy
(via Bj\"orken $x$) which depend on the interacting scale in a
medium. Therefore the nucleon rest energy is modelled by
introducing $x$-dependent effective nucleon rest energy (nucleon
mass) $M_x$ such that for large $x>0.6$ $M_x = {M_B} \simeq{M_N} +
\epsilon - 1.8 \bar{n}_A^{2/3}/M_B \approx {M_N-(20\div30)}$ MeV
(\ref{eq:mamam}) and for $x < 0.25$ it is equal to the free
nucleon mass $M_N$. In this way we have obtained a very good fit
to the experimental data \cite{Data} for $x > 0.15$ without
additional free parameters (contrary to claims made in \cite{MS}
where a similar approach but with only nucleon degrees of freedom
and constant mass was used). In our model only $\sim 1\%$ of
nucleon momenta is carried by additional partons from the sea
region, which is not in contradiction with lepton pair production
data \cite{Alde}. The medium modifications of mass proposed here
can be compared with the approach using the chiral soliton model
and direct quark-meson coupling (QMC) mechanism \cite{MSCH}. In
our approach we took effectively into account the nucleon-nucleon
interaction in NDIS and the final results are expressed by the
global nuclear parameters. The medium modifications depend on the
value of the average Fermi energy $E_{Fermi}$ and on the mass
defect $\epsilon$. We investigate the structure of the interacting
object and therefore its energy is not well define. In our
calculations of nucleon SF we try to take into account the
nucleon-nucleon interaction while we investigate its parton
structure.  Proposed changes of the nucleon rest energy should be
present in the nucleon structure obtained from the heavy ion
collision, however, this effect is of the same origin as the FSI
between hadrons (partons) and therefore it is rather subtle and
difficult to separate and detect. Nevertheless, some recent
observations of the decay spectrum of delta particles in high
energy heavy ion collision, seem to suggest that a similar
reduction ($\sim 20$ MeV) of its invariant mass exists \cite{MA}.
The MSR is satisfied within the conventional picture of
interacting nucleons, mainly due to the $x$-dependent effective
nucleon mass $M_x$ introduced.

Partial support of the Polish State Committee for Scientific
Research (KBN), grant 2P03B7522 (JR) and grants 621/E-78/SPUB/
CERN/P-03/DZ4/99 and 3P03B05724 (GW), is acknowledged.


\begin{thebibliography}{99}
\bibitem{Foot1} $M_N$ is nucleon mass, $x\!=\!x_{Bj}\!=\!Q^2/(2pq)$ partonic
                Bj\"orken variable and $Q^2\!=\!-q^2$. For fixed
                resolution, $Q^2\! =\! Q_0^2$, one has $x\! =\! x_{LC}\!
                =\!\frac{k^+}{p^+}$, where $x_{LC}$ is light cone nucleon
                rest frame variable with parton momenta $k$.
\bibitem{Foot2} The meaning of variables is the usual one: $x_{A}/A$
                is the ratio of the quark and the nucleus
                longitudinal momenta (the Bj\"orken variable for the
                nucleus), $y_{A}/A=p^{+}/P^{+}_{A}$ is the ratio of
                the nucleon and nuclear longitudinal momenta $p^{+}$,
                and $P^{+}_{A}$ and $x$ is the ratio of the quark and
                nucleon longitudinal momenta (the Bj\"orken variable
                for the nucleon). The function $\rho ^{A}(y_A)$ is
                the probability that a nucleon has a longitudinal
                momentum fraction $y_A$. Notice that momenta ratios
                $x$ and $y$ which occur here are defined on different
                energy scales: quark and nucleon, respectively.
\bibitem{Foot4} In fact $M_x$ depends also weakly on $Q^2$ because
                $M_x$ is constant in the sea region, $x<0.3$. We shall
                not discuss this possibility here. In addition note
                that our idea of $x$ dependent nucleon rest energy
                in medium does not affect nucleon mass in the equation
                of state for nuclear matter because the $x$ dependent
                longitudinal resolution $z$ is connected with the size
                of the virtual foton.
\bibitem{FOOTA} The values of parameters for the $Fe$ nucleus are:
                $M_N-M_B\simeq 27$ MeV and $p_F\simeq 1.2fm^{-1}$.
\bibitem{Foot6} In fact it could be even tuned to region $x<0.15$ (by
                introducing additional parameter, like pion rest
                energy) but for small $x$ the general assumptions of
                the convolution approach used here are not satisfied.
\bibitem{Foot7} The full analysis of the $A$-dependence with
                realistic nucleon distributions in nuclei will
                be presented elsewhere.
\end{thebibliography}
\end{document}